\documentclass[%
preprint,
 amsmath,amssymb,
 aps,
]{revtex4-2}

\usepackage{epstopdf} 

\usepackage{graphicx}

\usepackage{dcolumn}
\usepackage{bm}
\usepackage{hyperref}

\usepackage{epstopdf}
\usepackage{lineno}
\usepackage{textcomp}
\usepackage[gen]{eurosym}
\usepackage{slashed}
\usepackage[dvipsnames]{xcolor}  
\usepackage{placeins}

\usepackage{fancyheadings}
\usepackage{lastpage}
\usepackage{caption} 
\usepackage{verbatim}
\usepackage{color} 
\usepackage{subcaption} 
\usepackage{datetime}
\usepackage{booktabs, tabularx, multirow, array} 
\usepackage[separate-uncertainty,retain-explicit-plus,per-mode=symbol,binary-units]{siunitx}
\usepackage[version=4]{mhchem}
\DeclareSIUnit\c{\mbox{$c$}}
\DeclareSIUnit\year{yr}

\begin{document}

\preprint{APS/123-QED}

\title{Searching for ER and/or NR-like dark matter signals with the especially low background liquid helium TPCs}

\author{Junhui Liao}
\email{junhui\_liao@brown.edu}
\affiliation{%
 Division of Nuclear Physics, China Institute of Atomic Energy, Sanqiang Rd. 1, Fangshan district, Beijing, China, 102413. \\
Department of Physics, Brown University, Hope St. 182, Providence, Rhode Island, USA, 02912.
}%

\author{Yuanning Gao}
\email{yuanning.gao@pku.edu.cn}
\affiliation{%
 School of Physics, Peking University, Chengfu Rd. 209, Haidian district, Beijing, China, 10084. 
}%

\author{Jiangfeng Zhou, Zhuo Liang, Zhaohua Peng, Jian Zheng,  Haipeng An, Fengbo Gu, Meiyuenan Ma}
\affiliation{%
Division of Nuclear Physics , China Institute of Atomic Energy, Sanqiang Rd. 1, Fangshan district, Beijing, China, 102413. 
}%

\author{Lifeng Zhang, Lei Zhang}
\affiliation{%
Division of Nuclear Synthesis Technology, China Institute of Atomic Energy, Sanqiang Rd. 1, Fangshan district, Beijing, China, 102413. 
}%

\author{Xiuliang Zhao}
\affiliation{%
School of Nuclear Technology, University of South China, Shangsheng West Rd.28, Hengyang, Hunan, China, 421009. 
}%

\author{Junfeng Xia}
\affiliation{%
Shanghai Electric Cable Research Institute Co.,Ltd., Jungong Rd.1000, Yangpu District, Shanghai, China, 200093. 
}%

\author{Gang Liu, Shangmao Hu}
\affiliation{%
lectric Power Research Institute, CSG, Guangzhou 510663, China;\\
National Engineering Research Center of UHV Technology and Novel Electrical Equipment Basis,  Kunming 651705, China;
}%

\date{\today}

\begin{abstract}
In the Dark Matter (DM) direct detection community, the absence of convincing signals has become a ``new normal''  for decades. Among other possibilities, the ``new normal'' might indicate that DM-matter interactions could generate not only the hypothetical NR (Nuclear Recoil) events but also the ER (Electron Recoil) ones, which  have often been tagged as backgrounds historically. Further, we argue that ER and NR-like DM signals could co-exist in a DM detector's same dataset. So in total, there would be three scenarios we can search for DM signals: (i) ER excess only, (ii) NR excess only, and (iii) ER and NR excesses combined. To effectively identify any possible DM signal under the three scenarios, a DM detector should (a) have the minimum ER and NR backgrounds and (b) be capable of discriminating ER events from NR ones. Accordingly, we introduce the newly established project, ALETHEIA, which implements liquid helium-filled TPCs (Time Projection Chamber) in hunting for DM. Thanks to the nearly single-digit number of ER and NR backgrounds on 1 ton*yr exposure,  presumably, the ALETHEIA detectors should  be able to identify any form of DM-induced excess in its ROI (Research Of Interest). As far as we know, ALETHEIA is the first DM direct detection experiment claiming such an inclusive search;  conventional detectors search DM mainly on the ``ER excess only'' and/or the ``NR excess only'' channel, not the ``ER and NR excesses combined'' channel. 
\end{abstract}

\maketitle


\section{The ``new normal'' of dark matter direct detection} \label{sec1TheNewNormal}

Plenty of astronomical and cosmological observations have confirmed that Dark Matter (DM) exists in the Universe, the Milky Way, and the Solar System in which we live~\cite{Zwicky33, Rubin70, Refreiger03, Clowe06, Fields06, Planck2018ResultsOne, GaiaExp, HagenHelmi18, PostiHelmi19}. Particle physicists, however, know nothing about DM's particle characters. The Weakly Interactive Massive Particles (WIMPs) paradigm~\cite{SteigmanTurner85} is the most discussed one, possibly because it proposes a reasonable solution to interpret today's observed relic density of the universe with hypothetical particles sounding natural in particle physics~\cite{Feng10}. Unfortunately, evidence of convincing WIMPs signals has been missed for decades, though complementary searches have been employed in the satellites in the sky, colliders of several kinds on the ground, and low-background detectors in underground labs. We limit our discussions to underground direct detection in this paper. The ``new normal'' of null WIMPs signal has initiated many exotic ideas in identifying DM, theoretically and experimentally.

References~\cite{Feng10, BertoneHooperSilk2005} reviewed many particle DM candidates: WIMPs, superWIMPs, light gravitinos, hidden DM, sterile $\nu$, axions, supersymmetric candidates, light scalar dark matter, dark matter from little Higgs models, Kaluza-Klein states, superheavy dark matter, and Q-balls. In recent years, lots of new DM models/interpretations have arisen, including but not limited to the following categories: (a) Lower DM mass, all the way down to $\sim$ MeV/c$^2$ DM~\cite{FengKumar08, Feng10, Bohm13, Cohen93, Zurek14}. (b) New coupling or interaction, for instance, DM-electron coupling~\cite{Essig12}, Strongly Interacting Massive Particle (SIMP) models~\cite{StrasslerZurek07, Hochberg14, Hochberg15, Kuflik16, Kuflik17}, Effective Field Theory (EFT) operators~\cite{Fan2010, Fitzpatrick2013, Nikhil2014}, and DM absorption~\cite{Dror20}. (c) ``Exotic'' DM scenarios such as boosted dark matter, exothermic dark matter, and bosonic dark matter~\cite{Leane2022}. (d) Portable models~\cite{DiFranzo2013, Alves2014, Buckley2015}. (e) O-Helium (OHe) atoms, which is the composite of a hypothetical double-charged particle O$^{--}$ and a $^4$He atom, $^4$He$^{++}$ O$^{--}$~\cite{Khlopov21}. The ``Snowmass2021'' review paper~\cite{snowmass2021ParticleDMWP4} also introduced other interesting DM models.  We mainly discuss particle dark matter in the paper, though we will touch a little on axions in the following.

In the experimental community, LXe TPC experiments have pioneered progress in the high-mass WIMPs region ($\sim$ 10 GeV/c$^2$ - 10 TeV/c$^2$) in the past fifteen years. For instance, the backgrounds of the XENON experiment mitigated from XENON-10's 0.6 events/(kg $\cdot$ day $\cdot$ keV$_{\text{ee}}$)~\cite{XENON2008}  to XENONnT's 15.8$\pm$1.3 events/(ton $\cdot$ yr $\cdot$ keV$_{\text{nr}}$)~\cite{XENONnT-NR-2023} or 4.3$\times 10^{-5}$ events/(kg $\cdot$ day $\cdot$ keV$_{\text{nr}}$), which corresponds to a 4-order mitigation. XENON's upper limits have been improved from $\sim$ 10$^{-44}$ cm$^2$~\cite{XENON2008} to $\sim$ 10$^{-48}$ cm$^2$~\cite{XENONnT-NR-2023} from 2008 to 2023 thanks to the improved backgrounds, greater exposures, and other technical developments.   In the low-mass WIMPs region ($\sim$ 10 MeV/c$^2$ - 10 GeV/c$^2$), the progress is relatively slower.

Facing the fact of no convincing DM signals, particle physicists mainly figured out two searching strategies: (a) in the high-mass WIMPs region, exploring ``smaller cross-sections''  and (b) looking for  ``lighter DM'' signals in the low-mass area. The search for ``smaller cross-sections'' refers to exploring the cross-sections towards the atmospheric ``neutrino floor''~\cite{Billard14} or the ``neutrino fog''~\cite{OHare21}. Although the lowest 90\% CL (Confidence Level) upper limits for $\sim$ 30 GeV/c$^2$ WIMPs-nucleon interaction has already been set to be $\sim$10$^{-48}$ cm$^2$ by LXe TPC experiments~\cite{LZ2022, PandaX2021, XENONnT-NR-2023}, LXe experimentalists are planning to build a multi-ten tons LXe detector~\cite{NextGenerationLXeObservatory22}, which aims to reach the cross-section of $\sim$10$^{-49}$ cm$^2$, therefore, fully touching down the atmospheric neutrino floor/fog.
In the LAr community, DarkSide-20k would achieve a cross-section equivalent to LZ and XENON-nT. With 1000 ton*yr exposure, the Argo project would touch down the atmospheric neutrino floor/fog~\cite{DarkSide20k17}.

The search for ``lighter DM'' means looking for DM signals in sub-GeV/c$^2$ or even $\sim$ MeV/c$^2$ territory. Here, the ``lighter DM'' is not necessary to be WIMPs. Many exotic ideas with the ROI on the mass region were proposed (including our ALETHEIA project), as listed in reference~\cite{snowmass2021ParticleDM-lowThreshold}. The current upper limit for $\sim$ 100s MeV/c$^2$ - 10 GeV/c$^2$ is roughly between $\sim$10$^{-37}$ cm$^2$ and 10$^{-43}$ cm$^2$. Given that the $^8$B Solar neutrino floor/fog is roughly $\sim$10$^{-45}$ cm$^2$, there exist two to eight orders of cross-sections to be explored in the low-mass region before reaching the $^8$B neutrino floor/fog. 

No matter which DM models to test and which mass interval to look for, searching for a particle-like DM signal essentially is to discover the interesting signals registered in a DM detector among ``useless'' backgrounds. According to the traditional SI (Spin-Independent) model, the signals are single-scattering NR events, while the backgrounds are ERs and multiple-scattering NRs. 
Actually, ER could also be DM signals under certain scenarios. For instance, DM-electron coupling~\cite{Essig12}, the axioelectric effect produced by hypothetical solar axions in the Sun~\cite{Dimopoulos86}, and other models. Different from the QCD axions~\cite{PecceiQuinn77, Wilczek78, Weinberg78}, solar axions might be produced in the Sun through varying mechanisms~\cite{Primakoff51, Moriyama95, Redondo2013}. XENON collaboration observed ER events excess in their XENON-1T dataset~\cite{XENON1TERExcess2020}, though no similar excess showed up in the XENON-nT data~\cite{XENONnTERAnalysis2022}; PandaX-4T did not see noticeable ER excess neither~\cite{PandaX4T-ER2022}. 

Given that tens of elementary particles exist in the Standard Model (SM), it is reasonable to hypothesize that more than one type of ``dark elementary particle'' exists in the DM domain. If this is the case, it would be natural that some types of DM particles generate ER events in a detector through one unknown interaction while other types of DM register NR signals in the detector's same dataset via another different unknown interaction. Moreover, even if DM comprises WIMPs alone, it could also register both ER~\cite{Essig12} and NR signals. In summary, DM can leave ER \underline{and} NR events in a DM detector. 

Although nobody knows for sure which kind of recoil(s) DM will register in an underground detector,  from a detector's point of view, there might exist three kinds of excesses in the detector's dataset: (i) ER excess only, or (ii)  NR excess only, or (iii) ER and NR excesses combined. To capture possible DM signals under such a challenging scenario, ideally, a DM detector should (a) have the minimum (e.g. single-digit number) intrinsic ER and NR backgrounds and (b) be capable of discriminating the two kinds of recoils with high efficiency. This way, a single-digit number of ER and/or NR excessing events could trigger an observation (often is a 3 $\sigma$ significance) or even a discovery (corresponds to a 5 $\sigma$ significance)~\cite{DarkSide20k17}. On the other hand, imagining an underground DM detector has thousands of ER and/or NR backgrounds (which is the case for many currently running DM detectors), even if a small number of signal events existed in the dataset (hid in the bin with hundreds of events or the high density population), they would be buried in the backgrounds ``sea'' and mistakenly get rejected.

Taking the LZ detector as a template~\cite{LZTDR17}, our preliminary analysis showed that a three-year running with the 1.5-m size, 0.33-ton LHe (liquid helium) filled TPC would only have 11 ER and 0.5 NR background events, respectively. Such especially low ER and NR backgrounds make the detector alert to any possible excess caused by galactic DM particles, whether ER only, NR only, or both excesses come together. For details of the background estimation, please refer to our previous paper~\cite{ALETHEIA-EPJP-2023}. 

\section*{Summary}

Though the upper limits of WIMPs-nucleon interaction have been pushed down significantly in the past decades, no convincing signals have gradually become a ``new normal''. Historically, ER events have been treated as backgrounds by the mainstream of the DM direct detection community. Before convincing results are obtained, DM direct searches should not exclude ER as a possible singal completely, which means ER and NR could both be DM candidate events. Accordingly, there would exist three possible excesses in a detector's dataset: (a) ER only, (b) NR only, or (c) ER and NR co-existing. Under such scenarios, a DM detector should have minimum ER and NR backgrounds. Thanks to the nearly single-digit number of background events for both ER and NR with $\sim$ 1 ton*yr exposure, the ALETHEIA TPCs are well prepared for such searches.

\section*{Acknowledgments}


Junhui Liao would also thank the support of the ``Yuanzhang'' funding of CIAE to launch the ALETHEIA program. This work has also been supported by NSFC (National Natural Science Foundation of China) under the contract of 12ED232612001001.

\bibliography{tentativePRX}

\end{document}